\documentclass[sigconf]{acmart}
\usepackage[ruled,vlined]{algorithm2e}
\usepackage{amsmath}
\usepackage{silence}
\usepackage{balance}
\usepackage{flushend}
\usepackage{multicol}

\AtBeginDocument{%
  \providecommand\BibTeX{{%
    \normalfont B\kern-0.5em{\scshape i\kern-0.25em b}\kern-0.8em\TeX}}}
\setcopyright{acmcopyright}
\copyrightyear{2020}
\acmYear{2020}

\acmConference[SC '20]{SC '20: ACM/IEEE Supercomputing Conference}{November 17–19, 2020}{Online}
\acmBooktitle{SC '20: ACM/IEEE Supercomputing Conference,
  November 17–19, 2020, Online}
\acmPrice{15.00}
\acmISBN{978-1-4503-XXXX-X/18/06}

\begin{document}

\title{NetGraf: A Collaborative Network Monitoring Stack for Network Experimental Testbeds}

\author{Divneet Kaur}
\affiliation{\institution{Department of Electrical and Computer Engineering,\\ University of California San Diego,\\ San Diego, CA, USA}}
\email{dikaur@ucsd.edu}

\author{Bashir Mohammed}
\affiliation{\institution{Computing Research Division,\\ Lawrence Berkeley National Laboratory,\\ Berkeley, CA, USA}}
\email{bmohammed@lbl.gov}

\author{Mariam Kiran}
\affiliation{\institution{Scientific Networking Division,\\ Lawrence Berkeley National Laboratory,\\ Berkeley, CA, USA}}
\email{mkiran@es.net}

\begin{abstract}
Network performance monitoring collects 
heterogeneous data such as network flow data to give an overview of network performance, and other metrics, necessary for diagnosing and optimizing service quality. However, due to disparate and heterogeneity, to obtain metrics and visualize entire data from several devices, engineers have to log into multiple dashboards.

In this paper we present NetGraf, a complete end-to-end network monitoring stack, that uses open-source network monitoring tools 
and collects, aggregates, and visualizes network measurements on a single easy-to-use real-time Grafana dashboard. We develop a novel NetGraf architecture and can deploy it on any network testbed such as Chameleon Cloud by single easy-to-use script for a full view of network performance in one dashboard.

This paper contributes
to the theme of automating open-source network monitoring tools software setups and their usability for researchers looking to
deploy an end-to-end monitoring stack on their own testbeds.

\end{abstract}

\keywords{Network monitoring tools, real-time dashboards, deployable solution}

\maketitle

\section{Introduction}
Cloud  infrastructure   is   composed   of   heterogeneous resources made up of hardware, virtualization, storage, and networking components \cite{marinescu2017cloud}. Network performance monitoring  (NPM)  is  the  process  of  visualizing,  monitoring,  optimizing, troubleshooting and reporting on the service quality of your network as experienced by your users \cite{narayana2017language}. 
Different NPM tools collect different data such as packet loss, network flow which when combined provides a complete picture of the network infrastructure. This helps monitor and analyze a network's performance, availability, and  other  important  metrics.
But this  heterogeneity  comes  with  a  challenge  when
network engineers try to visualize all the  network metrics coming  from  different monitoring  tools  without single  dashboard.  


 

We develop NetGraf, a collaborative network monitoring stack that provides a holistic view of the network system providing visualizations of various metrics from different monitoring tools in a single dashboard for valuable insight on the network. While developing monitoring stack, our main contributions are:
\begin{itemize}
	\item We explore six diverse open-source network monitoring tools and package them for any network infrastructure by developing a monitoring pipeline that fetches data from these tools, aggregates them and stores in a time-series database \cite{taherizadeh2016runtime}.
    \item We develop an Application Programming Interface (API) to define interactions between these tools and Grafana, an open-source visualization software, in order to generate visualizations of collected metrics and network statistics.
\end{itemize}

\balance

\begin{table*}[htbp]
\caption{Chameleon Core Network Environment settings.}
\begin{center}
\begin{tabular}{|c|c|c|c|c|c|c|c|c|}
\hline
\textbf{Node}&\multicolumn{6}{|c|}{\textbf{Network Monitoring Tools on Chameleon testbed}} \\
\cline{2-7} 
\textbf{Specification} & \textbf{{Prometheus}}& \textbf{{ntopng}}& \textbf{{netdata}}& \textbf{{perfSONAR}}& \textbf{{Zabbix}} & \textbf{{Grafana}}\\
\hline
IP Address & 192.168.100.11& Installed on all nodes &Installed on all nodes&192.168.100.16&192.168.100.13&192.168.100.14\\
\hline
Floating IP & 192.5.87.178& Installed on all nodes & Installed on all nodes& 192.5.87.157 & 192.5.87.126  & 192.5.87.126\\
\hline
Gateway  & 192.168.100.1 &192.168.100.1  & 192.168.100.1  &192.168.100.1  & 192.168.100.1 & 192.168.100.1  \\
\hline
Listening Port &9090& 3000 & 19999& 861& 10050& 3000  \\
\hline
Switches  & Corsa Switch&  Corsa Switch&  Corsa Switch&  Corsa Switch&  Corsa Switch &  Corsa Switch \\
\hline
OS & CC-Ubuntu18.04&CC-Ubuntu18.04/16.04,  &  CC-Ubuntu18.04/16.04,&CC-Ubuntu18.04& CC-CentOS7 &CC-Ubuntu18.04 \\
 & &Centos7  &  Centos7&& &\\
\hline
\end{tabular}
\label{table1}
\end{center}
\end{table*}

\begin{figure*}[htbp]
\centering
    \includegraphics[width=5in]{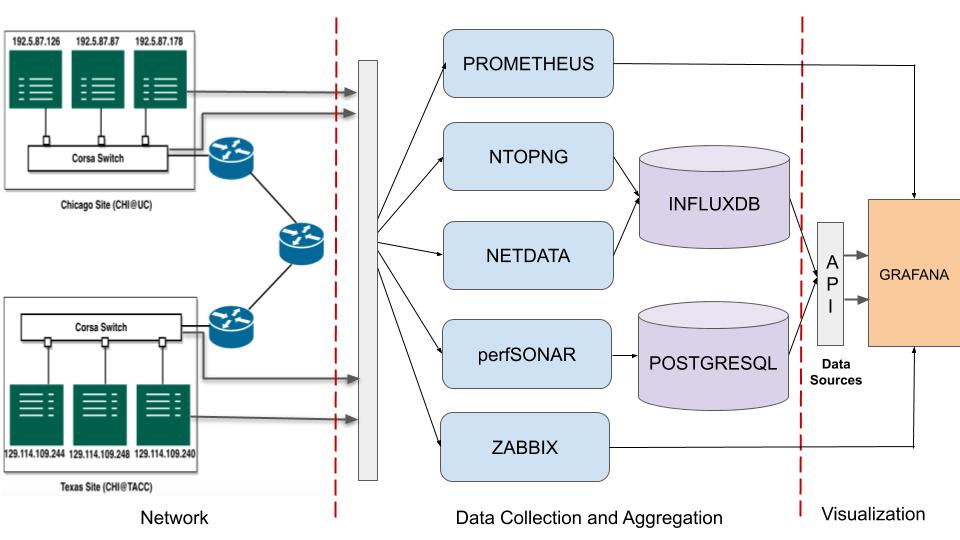}        \caption{NetGraf architecture composes of three main modules: Network, Data collection and Aggregation, Visualization module.} 
         \label{fig:fulltop}
\end{figure*}

\balance

\begin{figure*}[htbp]
\centering
    \includegraphics[width=5in]{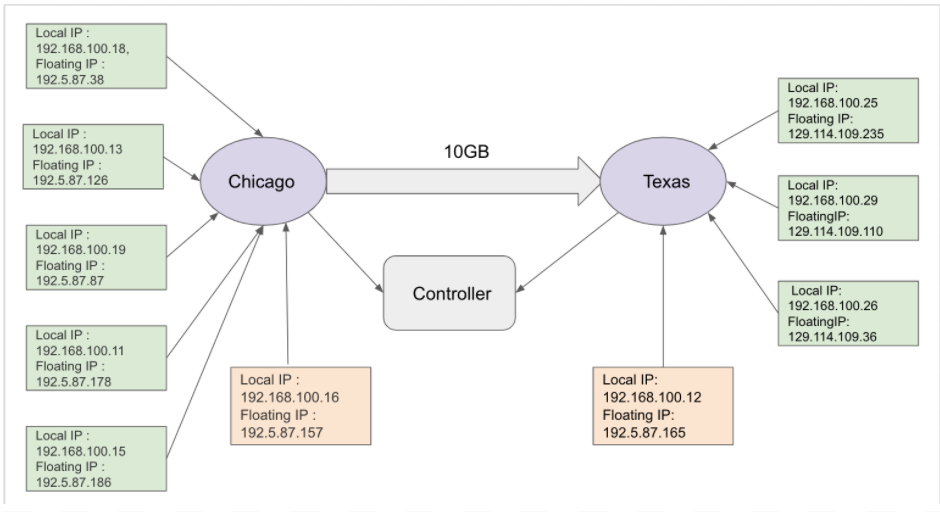}
     \caption{NetGraf showing the network topology deployed on the Chameleon testbed.}
     \label{fig:plot12}
\end{figure*}

\section{Methodology: NetGraf Architecture}

Figure \ref{fig:fulltop} presents the NetGraf architecture. 
NetGraf is designed for supporting multiple network monitoring tools, for different network devices such as large scale hardware infrastructure and application servers. It supports seamless plug-in for new NPM tools. It also allows monitoring, collecting, identifying and visualizing a wide spectrum of network data in a single dashboard that enables network engineers to identify   mishaps like degradation and outages that occur in a network experimental testbed.


The architecture consists of three main modules: 

\subsubsection{Network and Application Module} The network topology deployed on Chameleon testbed with monitoring tools. (Figure 2)

\balance
\begin{figure*}[htbp]
\centering
    \includegraphics[width=6in]{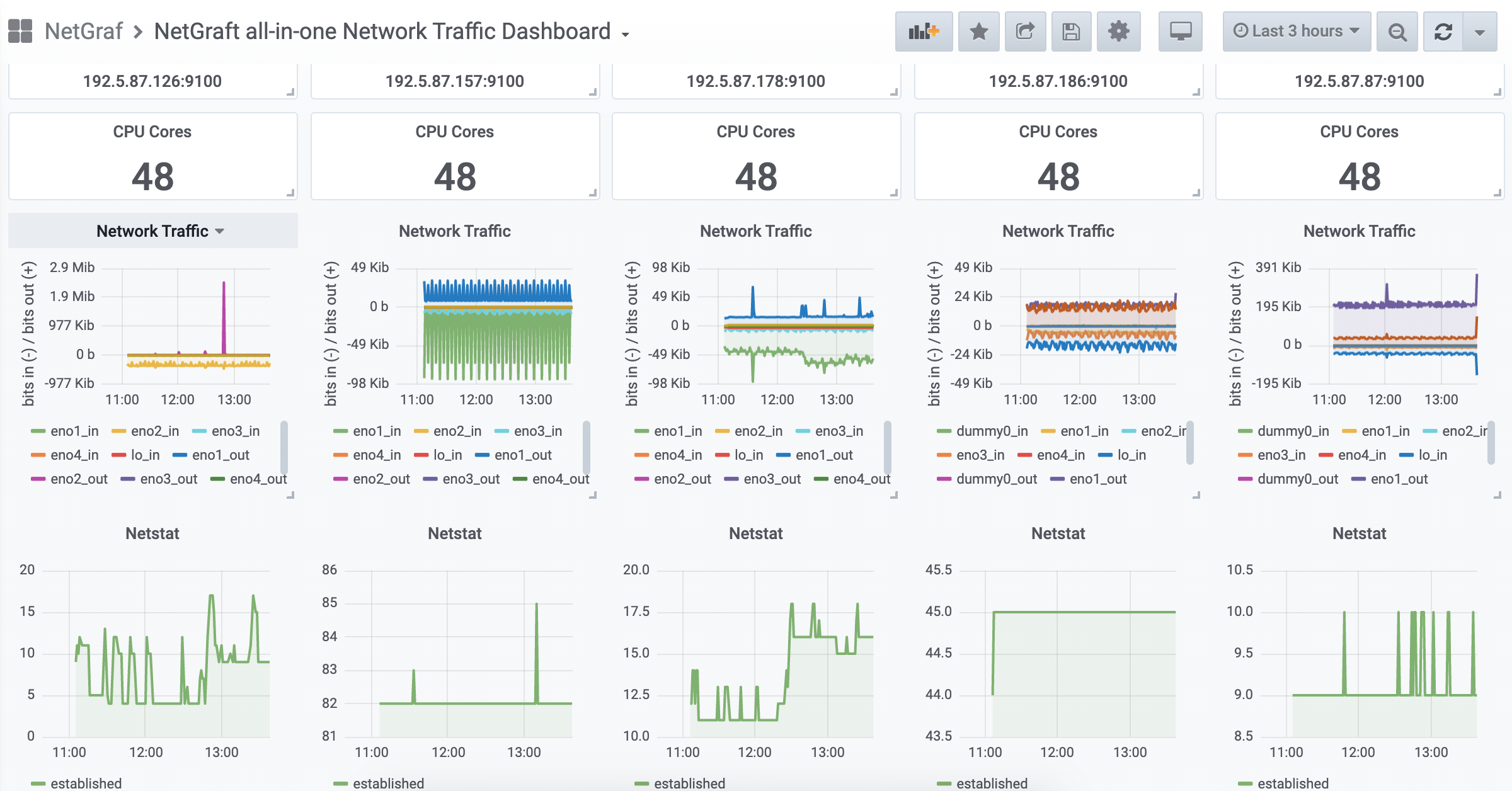}
     \caption{NetGraf showing a snapshot of network metrics on Grafana.}
     \label{fig:plot2}
\end{figure*}

\subsubsection{Data Collector and Aggregator Module}
Various monitoring tools were installed to collect metrics and network statistics, such as ntopng and netdata (Table \ref{table1}). Since Prometheus can scrape metrics from multiple nodes, it is installed on one node and scrapes metrics from all other nodes. 
Zabbix was installed on one node collecting server related metrics. perfSONAR, a network measurement toolkit, was installed on Chicago and Texas sites.

These tools were connected to databases for storage - ntopng, netdata were connected to InfluxDB, a time-series database. Prometheus, Zabbix have an inbuilt database. perfSONAR's collected results were archived in a relational database, postgreSQL. 

\subsubsection{Monitoring and Visualization Module}
To generate visualizations from the available metrics, we created an API between the databases and Grafana. This was established by adding Influxdb, Postgresql, Prometheus and Zabbix as \emph{data sources} in Grafana. 

Since a large number of metrics were being collected, we performed an elimination process where network metrics such as Transmission Control Protocol, throughput and loss were selected. This helped us create a dashboard with relevant metrics only.


\section{Results: Lessons Learnt}

Figure \ref{fig:plot2} shows a snapshot of five nodes from Chameleon testbed located at Chicago for 3 hours. 

While identifying efficient routes to connect the monitoring tools to Grafana, we explored two other approaches,
\begin{itemize}
\item The monitoring tools were connected to Prometheus which recorded only node metrics such as disk storage, which was not network data.
\item The data was directly fed to Grafana using plugins. Due to lack of direct plugins for all tools and databases for storage, this approach was not efficient as well.
\end{itemize}

\section{Conclusion and Future Work}
 
Monitoring and understanding network infrastructure performance is essential to learn network experimentation. In this work, we present a unique monitoring approach which can collect and store network metrics and solve the heterogeneity of diverse network monitoring tools by displaying them all in a single dashboard. We have created two users - admin and viewer to allow many people to view the dashboard
In future, we will apply machine learning algorithms to the dashboard to provide more insights on network performance and availability analysis.

\balance
\begin{acks}

We would like to thank Paul Ruth for the technical support. This work was supported by U.S. Department of Energy, Office of Science Early Career Research Program for `Large-scale Deep Learning for Intelligent Networks' Contract no  FP00006145.
\end{acks}

\balance
\bibliographystyle{ACM-Reference-Format}
\bibliography{biblio}

\end{document}